\documentclass[conference]{IEEEtran}
\usepackage{cite}

\ifCLASSINFOpdf
\else
\fi
\usepackage{todonotes}
\usepackage{tikz}
\usepackage{amsmath}
\usepackage{cleveref}

\usepackage{url}
\usepackage{float}
\usepackage{xurl}
\usepackage{wrapfig}
\usepackage{tablefootnote}
\usepackage{fontawesome}
\usepackage{graphicx}

\usepackage{subfig}

\usepackage{xltabular}
\usepackage{tabularx}
\usepackage{changepage}
\usetikzlibrary{tikzmark}
\usepackage{booktabs}
\usepackage{multirow}
\usepackage{makecell}
\usepackage{multicol}
\usepackage{threeparttable}
\usepackage{array}

\usepackage{listings}
\usepackage{xcolor}

\lstdefinestyle{highlight}{
    basicstyle=\ttfamily\small\color{red}, 
}

\begin{document}
%
\title{From Blocking to Breaking: Evaluating the Impact of Adblockers on Web Usability}

\author{\IEEEauthorblockN{Ritik Roongta$^*$}
\IEEEauthorblockA{New York University\\
ritik.r@nyu.edu}
\and
\IEEEauthorblockN{Mitchell Zhou$^*$}
\IEEEauthorblockA{New York University\\
mz2909@nyu.edu}
\and
\IEEEauthorblockN{Ben Stock}
\IEEEauthorblockA{CISPA Helmholtz Center \\for Information Security\\
stock@cispa.de}
\and
\IEEEauthorblockN{Rachel Greenstadt}
\IEEEauthorblockA{New York University\\
greenstadt@nyu.edu}}

\maketitle
\def\thefootnote{*}\footnotetext{These authors contributed equally to this work}

\begin{abstract}
Recent years have seen a sharp rise in adblocker use, driven by increased web tracking and personalized ads. However, a significant issue for adblocker users is the web breakages they encounter, which worsens their browsing experience and often leads them to turn off their adblockers. Despite efforts by filter list maintainers to create rules that minimize these breakages, they remain a common issue. Our research aims to assess the extent of web breakages caused by adblocking on live sites using automated tools, attempting to establish a baseline for these disruptions.
The study also outlines the challenges and limitations encountered when measuring web breakages in real-time. The current automated crawler's inability to consistently navigate a vast array of websites, combined with the unpredictable nature of web content, makes this research particularly difficult. We have identified several key findings related to web breakages in our preliminary study, which we intend to delve deeper into in future research.
\end{abstract}


%

\section{Introduction}
\label{sec:introduction}

In the ever-evolving landscape of the digital age, online advertising has become increasingly intrusive. Adblockers, once a niche tool for the tech-savvy, have expanded into mainstream usage, signaling a significant shift in how users interact with online content. According to a recent report by Pagefair~\cite{blockthrough}, the global number of internet users who block ads surpassed 820 million in 2022. Additionally, a study revealed that 22.2\% of all internet users and 30\% of those using Chrome have installed AdBlock Plus~\cite{pujol15imc}. In a 2022 advisory, the FBI recommended that individuals and companies consider using adblockers for enhanced online security and privacy~\cite{fbi-adblock}.

\textbf{Web breakage} describes situations where a website does not work as expected, leading to issues such as malfunctioning pages, incorrectly displayed content, or features not operating correctly. These situations at times can arise due to the website's inherent functionality towards adblockers such as `disable adblocker` prompts. These breakages can be either visible, such as easily noticeable issues with site layout or functionality, or invisible, manifesting in less direct ways like diminished content quality, intrusive low-quality ads, or extended website load times. While adblockers contribute to a more secure and private browsing experience, they also modify how a website's elements load and interact and can inadvertently cause web breakages. Web developers may not always anticipate or account for these adblocker interactions, leading to unintended site malfunctions. Additionally, some web developers deliberately introduce disruptions to penalize adblocker users, potentially degrading their browsing experience.

Various researchers have approached the challenge of measuring web breakages indirectly, using different proxies instead of directly replicating the breakages themselves. A few of the notable examples include using functional JavaScripts to estimate web breakages by NoScript and Ublock Origin~\cite{amjad}, creating an ML classifier to predict potential breakages caused by specific filter list rules~\cite{smith}, and evaluating the potential for visible breakages by observing changes in the number of images and text on websites after installing adblockers~\cite{hieu}. A notable gap in these studies is the absence of comprehensive, large-scale evaluations of reproducible web breakages. For detecting adblocking, some researchers have identified the use of anti-adblocking scripts on websites as an indicator of a site's capability to detect adblockers~\cite{zhu18ndss, nithyanand16foci, iqbal17imc}, highlighting the fact that a large number of websites actively detect adblockers on their webpages.

Our work aims to address this by developing a measurement method that is robust against the inherent randomness of the web, allowing us to assess web breakages on live sites in real time. Our objectives are twofold: \textbf{first}, to establish a lower bound for the web breakages that can be identified across the internet; and \textbf{second}, to devise a system for automatically detecting web breakages through web crawling. This approach not only fills a crucial research gap but also enables us to analyze how the web's resilience to adblocking has evolved over several years.


We concentrate on \textbf{five} specific categories of breakages to highlight the occurrence of visible disruptions. They include detection of the extension by websites, malfunctioning HTML elements, failure to load resources, browser crashes, and pages being unresponsive or loading slowly. These categories are developed from the breakages mentioned in user reviews on the Chrome Web Store and existing databases documenting potential visible breakages~\cite{nisenoff, smith, adguard_filters}. Together, these categories represent approximately 70\% of all breakages mentioned on these forums.

To evaluate these web breakages, we utilize three distinct adblockers: Ublock Origin (UbO), Adblock Plus (ABP), and Privacy Badger (PB)~\cite{abp, ubo, pb}. These extensions were selected to encompass the various adblocking approaches used online. UbO is noted for its comprehensive and diverse filter lists for ad and tracker blocking. ABP distinguishes itself by including an "Acceptable Ads" list, which affects its impact on web breakages differently compared to other adblockers. Privacy Badger, on the other hand, uses a behavior-based blocking strategy rather than relying on filter lists, leading to variable effects on web elements under different conditions. 

We found that the actual number of websites that break in the presence of adblockers is quite small compared to the perceived notion. While  1.67\% of the websites show users a `disable adblocker' prompt in the case of ABP, just 0.67\% of the websites have breakages in HTML elements. We do find a significant number of websites that have missing resources but they do not always reflect as visible breakages. Page crashes and unresponsiveness are the least common breakages with 0 and 0.4\% websites falling into these categories respectively. We also discovered the limitations in the state-of-the-art web crawlers to conduct large-scale measurements and how websites respond to automated crawling hence affecting measurements.

\section{Background}
A browser extension or browser plugin is a small software application that enhances the capacity or functionality of a web browser. A browser extension leverages the same Application Program Interfaces (APIs) available to the website’s JavaScript, in addition to its own set of APIs, thus offering additional capabilities. Privacy-preserving extensions offer advanced functionality and use sensitive permissions, making them an attractive target for adversaries. Previous studies have extensively examined the privacy and performance of a subset of these extensions using static analysis and web measurement. Our research focuses on the breakage aspect of these extensions.

Nisenoff et al.~\cite{nisenoff} studied the GitHub issue reports and the Chrome web store user reviews for adblocking extensions to build a taxonomy for breakages experienced by users. They find seven broad categories to classify the breakages --- Loading and responsiveness, Resources and third-party content, Extension detection and interaction, HTML Elements, Browser level, Authentication and sessions, and Vague. They also mention subcategories within each broad category to understand the specific domains causing the breakages within each broad category. We use their taxonomy to support our methodologies as we measure the most important components of a webpage that users might consider being broken. Note that since authentication cannot be automated at scale, we omit this category as well as the all subcategories of \emph{Vague}.

Catapult is a tool developed by Chromium developers for performance analysis, testing, and monitoring of Chrome. They can also be used for analyzing and monitoring websites, and eventually Android apps. According to its official documentation~\cite{catapult}, it is a home for several performance tools that span from gathering, displaying, and analyzing performance data. This includes --- Trace-viewer, Telemetry, Performance Dashboard, Systrace, and Web Page Replay. In this work, we use the Web Page Replay tool to record the web pages and replay them deterministically to create a reproducible dataset of website breakages and enable multiple fetches of websites in short intervals of time to get a fresh state for each measurement. The latter is blocked by websites to prevent DDoS~\cite{ddos} attacks wherein a malicious actor can deploy botnets to imitate similar behavior. 

HTML serves as the foundational language for crafting and styling web pages and structuring content with elements such as text and images. XPath, a query language, facilitates the accurate identification and selection of these elements within HTML documents. Selenium~\cite{selenium}, an automation tool for web applications, offers a range of methods for locating, navigating to, and interacting with different web elements. The consistent nature of an element's XPath across multiple visits enables reliable interaction with and testing of various web components in a deterministic manner.

\section{Methodology}

In this section, we outline our approach for evaluating web breakages across various categories. Our methodology aims to fulfill two main goals. First, it's crucial to analyze web behavior across both highly popular as well as less frequented websites to grasp a wide range of experiences. This analysis will allow us to understand the relationship between a website's popularity and the prevalence of web breakages. Second, our tool must counter server-side variables, such as temporal cookies and timestamps, which introduce unpredictability into our measurements. We have designed our methodology to be resilient, ensuring these elements of randomness do not skew our findings. Roth et al.~\cite{roth} investigated how top websites security policies vary with different client settings, demonstrating that website responses can indeed differ based on client configurations.

To tackle the first challenge, we employ Selenium to navigate through a set of 20,000 websites. This set is curated by combining the top 10,000 websites with those ranked between 90,000 and 100,000 from the Tranco list~\cite{tranco}, allowing us to examine the behaviors of both highly and less frequently visited sites. We refine this list by excluding Content Delivery Networks (CDNs) and sites that are unreachable, specifically, those that result in a connection error or return a status code different from 200, culminating in a final pool of 16,790 websites. We then proceed to visit the homepages of these selected sites, forming our website testing pool. All website visits are conducted using the Chrome browser, recognized as the most widely used browser according to recent statistics~\cite{chrome}.


To minimize the effects of random content variations, we implement the Catapult tool in conjunction with Google Chrome. Catapult enables us to capture and later replicate the exact state of websites, using a precise set of resources for playback. It operates by setting up a proxy to download all resources requested by a website and then uses this same proxy along with the downloaded content to accurately re-render the websites.

We conducted a manual evaluation of Catapult's effectiveness on a randomly selected sample of 200 websites. During this process, we identified certain limitations with the tool (detailed in Section \ref{sec:appendix}). These findings not only highlight the challenges in creating a consistent web browsing environment but also contribute to the development of a reproducible dataset of web breakages for future validation and study. Since we did not find the catapult tool to be as efficient in its crude form, we did not use it for the preliminary study. We plan to enhance the catapult's effectiveness and then incorporate it into our future work.



Now we discuss the individual methodologies deployed for measuring breakages in each category as mentioned in \Cref{sec:introduction}.

\subsection{Extension detection}
\begin{figure}[htbp]
  \centering
  \includegraphics[width=0.9\columnwidth]{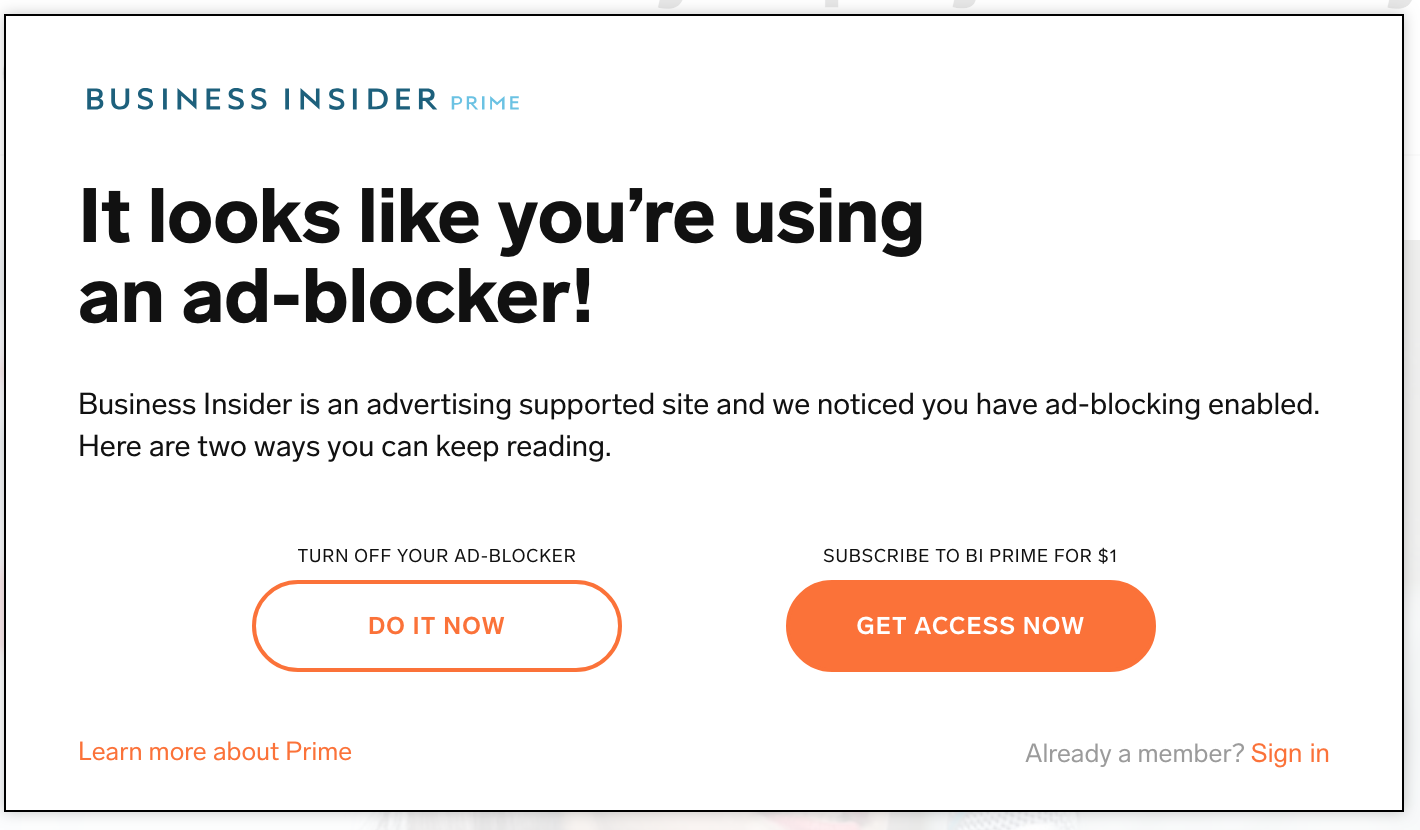}
  \caption{Examples of different `disable adblocker' prompts shn by various websites.}
  \label{fig:img3}
\end{figure}

A large number of websites detect adblocking extensions, as highlighted in the prior work by Amjad et al.~\cite{amjad}. Our preliminary studies show that only a fraction of them prompt users with a `disable adblocker` prompt. A `disable adblocker` prompt usually shadows the entire website and renders it useless unless the extension is turned off. Figure \ref{fig:img3} shows different kinds of prompts on various websites. We ran this measurement on 1,500 websites from the website testing pool to get preliminary insights into this category.

\begin{figure}[htbp]
  \centering
  \includegraphics[width=0.45\columnwidth]{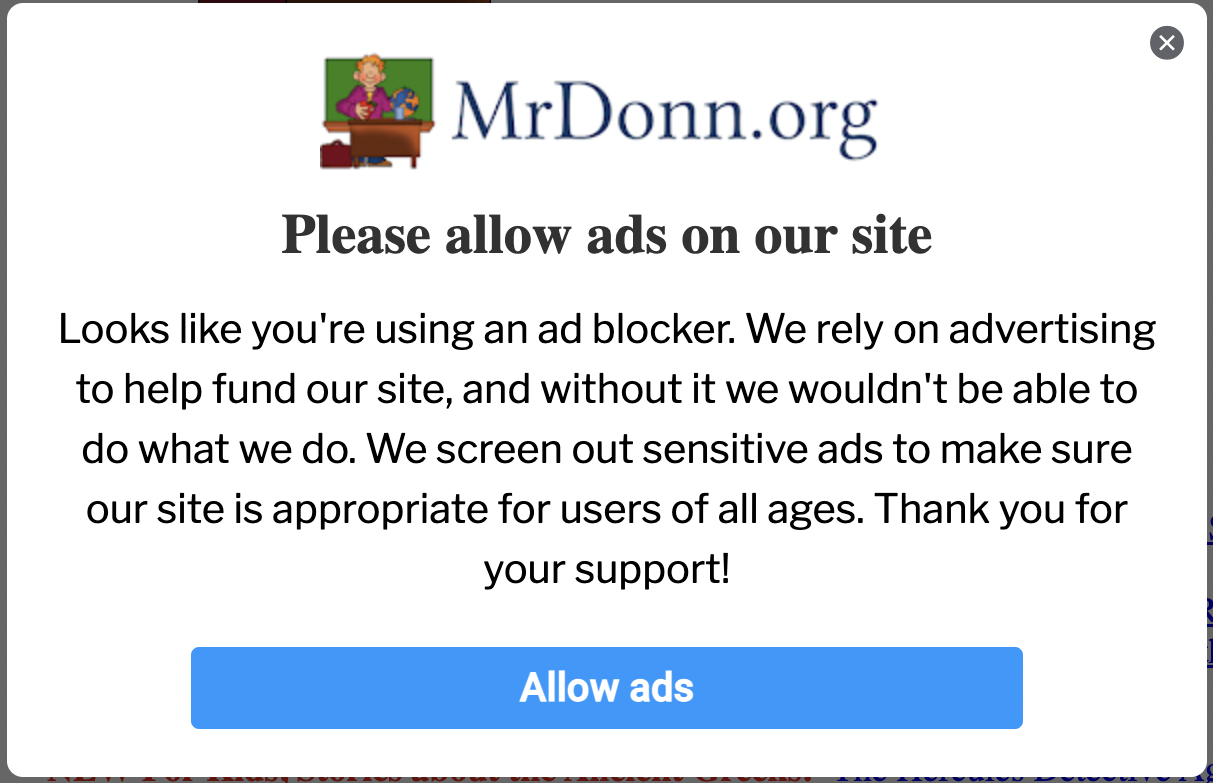}
  \hfill
  \includegraphics[width=0.45\columnwidth]{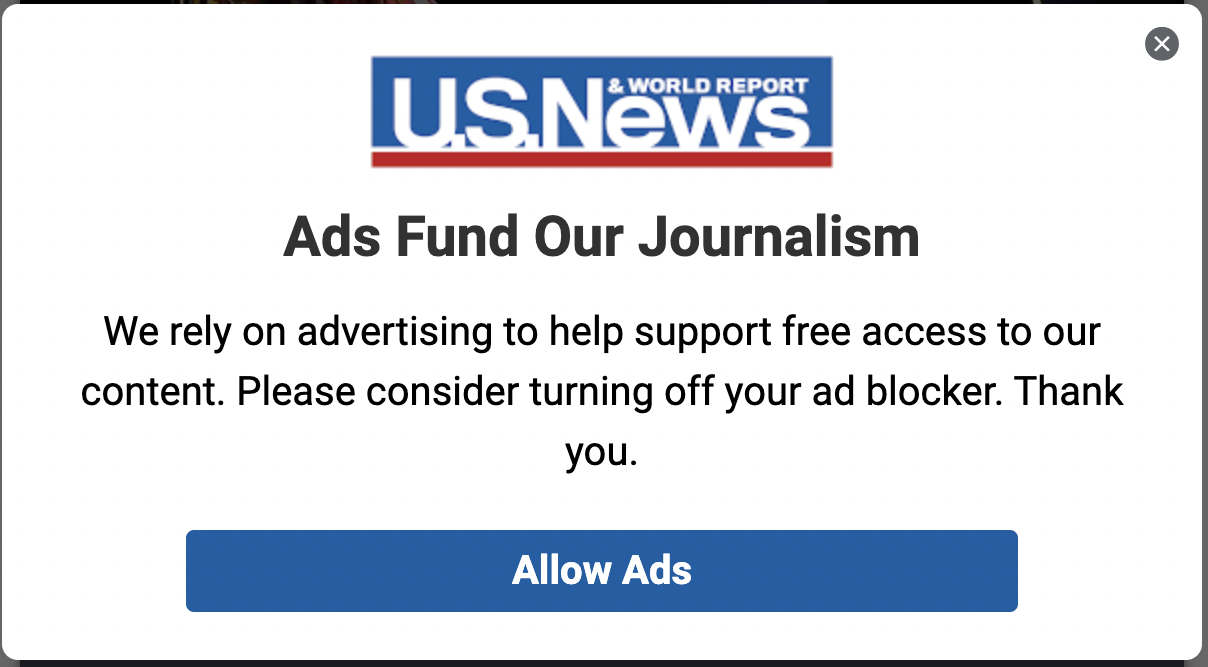}
\end{figure}

To investigate the presence of prompts urging users to disable adblockers, we employed a Selenium crawler operating in virtual display mode. This crawler navigated not only the landing page of each website but also three additional pages with the longest same-origin URLs. These longer URLs often lead to inner pages, as opposed to the shorter URLs, which typically link to different sections of the same landing page. We hypothesized that inner pages, likely to have higher content value and traffic, might be more aggressively protected against ad filtering by website administrators.

Our methodology involved visiting these websites and their inner pages, both with adblocking extensions enabled and without them. We examined the source code of every frame on these pages for specific keywords associated with adblocker detection, such as `adblocker', `detect', and `disable'. These keywords were chosen based on their common use in scripts or dialog boxes that ask users to disable their adblockers. The rationale behind focusing on these terms lies in their direct association with the mechanisms of adblock detection and the typical actions requested from users in response to detection.

Upon identifying a frame that contained any of these keywords, we tallied the website as one actively seeking to detect and possibly inhibit adblocker usage. To strengthen the reliability of our findings, we also captured screenshots of the identified prompts and conducted manual follow-up visits to these websites. This two-pronged approach of automated detection followed by manual verification allowed us to accurately identify and report on websites implementing measures to prompt users into disabling their adblockers.


\subsection{HTML elements}
An average website contains a large number of HTML elements. Usually, adblockers can result in breaking the functionality of these elements and rendering them useless. Almost 9\% of the threads report breakages in one of the other HTML elements. We concentrate on four HTML elements that are most frequently reported for causing issues. This ranking has been developed from Nisenoff et al.'s taxonomy of breakages which they build up from user reviews. They are login elements, buttons, links, and dropdown menus. Form-based HTML elements have been omitted as they are more complicated for automated interaction. We run this experiment on 1000 sites. After removing failed renders and sites with no HTML elements in the respective categories above, we are left with 300 sites for each category.

\begin{lstlisting}[caption={DOM tree before interaction with the button element on www.reddit.com}, label=lst:html]
<div>
<faceplate-expandable-section-helper rpl="">
<details class="p-0 m-0 bg-transparent border-none rounded-none">
<<< <summary aria-controls="Gaming" aria-expanded="false" class="font-normal">
<left-nav-topic-tracker noun="topic_menu" topic="gaming">
<li rpl="" class="relative list-none mt-0 " role="presentation">
</div>
\end{lstlisting}

\begin{lstlisting}[moredelim={[is][\color{blue}\ttfamily\small]{@@}{@@}}, caption={DOM tree after interaction with the button element on www.reddit.com. The changes are reflected in blue color.}]
<div>
<faceplate-expandable-section-helper rpl="">
<details class="p-0 m-0 bg-transparent border-none rounded-none">
@@>>> <summary aria-controls="Gaming" aria-expanded="true" class="font-normal"> @@      
<left-nav-topic-tracker noun="topic_menu" topic="gaming">
<li rpl="" class="relative list-none mt-0 " role="presentation">
</div>
\end{lstlisting}

To detect web breakages in the web elements caused by adblockers, we start by crawling websites without extensions to set a control case. This step records how web elements originally functioned. During this crawl, we select elements for interaction and record their Document Object Model (DOM) states, including the state of elements up to four levels in the DOM tree. Our quantitative analysis for the changes in the DOM tree reveals that in 85\% of the cases, DOM changes appear in levels with three nodes.  This approach helps to capture relevant changes without including unrelated alterations. The next step involves crawling the same websites with extensions enabled and interacting with the same elements. If there's no change in the DOM or the elements disappear when extensions are enabled, it's marked as `breakage'. This method involves comparing the DOM state before and after interactions to identify any discrepancies. Since a website can have many links and buttons, we randomly select at most 15 elements in each category for our measurement, ensuring a broad and representative assessment of the extension's impact. Listing 1 shows the DOM tree before interaction and Listing 2 shows the tree after interaction. Tha change is highlighted in line 4 with a blue color.

\subsection{Resources not loaded}
\begin{figure}[htbp]
  \centering
  \includegraphics[width=0.9\columnwidth]{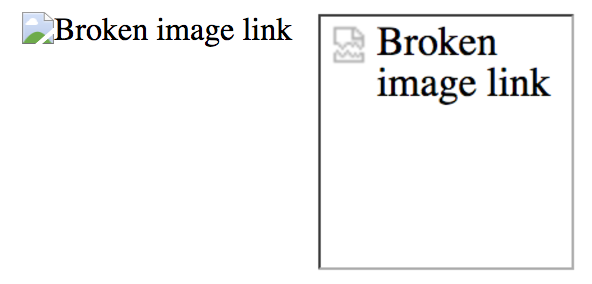}
  \caption{Examples of visual breakage when the image resource is not present or fetched by the browser.}
  \label{fig:img4}
\end{figure}

The most common and heavy resources loaded by websites are videos, images, and JavaScript files. The inability to load these resources can lead to data corruption and end up breaking the websites. To measure the number of resources not loaded, we use Browsermob proxy~\cite{browsermob} along with selenium. This proxy places itself between the Chrome instance and the external web and helps us capture the resources that are requested by the website. We capture the resources in both the control as well as the extension case and compare them to find resources that were not loaded when the adblocker was active. To get maximum coverage and counter the lazy loading of resources, we scroll to the bottom of the webpage every time we visit the website. We run this measurement on 1,500 sites and are left with 600 sites after deploying the filtering strategy discussed below. Figure \ref{fig:img4} shows a few examples of how breakages appear on the webpage if the image resource is not present.

To focus on specific resources such as videos, images, and JavaScript, our methodology involves initially filtering resources that remain consistent across crawls with and without adblocker extensions. Subsequently, we eliminate resources linked to ads or redirections by excluding those with non-200 status codes or URLs listed in widely used filterlists like Easylist, Easyprivacy, Fanboy, and Peter Lowe~\cite{peterlowe, easylist, filterlist}. This step ensures the removal of non-relevant resources. The remaining resources, absent in the extension-enabled crawls, are then revisited via the crawler. We take screenshots for visual resources, except for JavaScript elements, which are noted for future analysis as their screenshots contain limited meaning. We aim to address them in our future work involving code coverage. The process of taking screenshots helps us to discard resources with session-specific or dynamic URLs that can be attributed to third-party sources or ads. It further helps us to confirm the presence of local/static web resources affecting website functionality.

\subsection{Browser or page crashes}
Browser crashes or page crashes are very rare as the rules causing such breakages are easily detectable and hence patched in the filterlists that adblockers use. However, since users report instances of these kinds of breakages, we measure them using the logging functionality of Google Chrome. Selenium provides us with a `goog:loggingPrefs' hook to capture and study the entire Chrome logs for a website. If the log has a 404 status for the landing page or contains the `page crash/browser crash' keyword, we manually visit the websites and confirm the crashes. We run this experiment on 1000 websites, randomly chosen from the website testing pool.

\subsection{Page unresponsiveness}
This breakage category addresses the issue wherein a page might be very slow to load or become unresponsive in the presence of adblockers. Although it is easy to detect the website load time and unresponsiveness, attributing it to the adblockers is hard since the delay in page response is usually carried out at the server end. Since we do not have access to the website's servers, it is difficult to attribute the delay to adblockers based on the analysis of data on the client side. To identify if the website is unresponsive in the presence of extensions, we load it without the extension, with the extensions, and without the extension again. Each case is repeated five times. The last crawl is to identify if the websites maintain a persistent state of the users even after they remove adblockers. We run this experiment on 1,000 websites as well.

\begin{table*}[ht]
\centering
 \caption{Number of websites that break under different categories for each extension. The percentage of the total websites is shown alongside the data. The categories are highlighted on the top --- Extension Detection, Broken HTML Elements, Resources not loading, Page Crash, Page unresponsiveness. The number of websites used to collect data for each category is shown in brackets. The subcategories within each category are highlighted in \textit{italics}.}
\label{tab:data}
\begin{tabular}{!{\vrule width 1.5pt}l!{\vrule width 1.5pt}l!{\vrule width 1.5pt}lllll!{\vrule width 1.5pt}ll!{\vrule width 1.5pt}l!{\vrule width 1.5pt}l!{\vrule width 1.5pt}}
\multicolumn{1}{c}{\textbf{Categories}} & \multicolumn{1}{c}{\textbf{\begin{tabular}[c]{@{}c@{}}Detection\\\textit{(1500)}\end{tabular}}} & \multicolumn{5}{c}{\textbf{\begin{tabular}[c]{@{}c@{}}HTML Elements\\\textit{(300)}\end{tabular}}} & \multicolumn{2}{c}{\textbf{\begin{tabular}[c]{@{}c@{}}Resources\\\textit{(600)}\end{tabular}}} & \multicolumn{1}{c}{\textbf{\begin{tabular}[c]{@{}c@{}}Crash\\\textit{(1000)}\end{tabular}}} & \multicolumn{1}{c}{\textbf{\begin{tabular}[c]{@{}c@{}}Unresponsiveness\\\textit{(1000)}\end{tabular}}}\\
\midrule[1pt]
\textbf{Extension} & \rotatebox{90}{\textit{Disable Prompt}} & \rotatebox{90}{\textit{Buttons}} & \rotatebox{90}{\textit{Links}} & \rotatebox{90}{\textit{Login}} & \rotatebox{90}{\textit{Drop Downs}} & \rotatebox{90}{\textit{Input}} & \rotatebox{90}{\textit{Images}} & \rotatebox{90}{\textit{Videos}} & \rotatebox{90}{\textit{Page Crashes}} & \rotatebox{90}{\textit{Count}} \\ 
\midrule[1pt]
ABP & 25 $^{1.67\%}$ & 1 $^{0.33\%}$ & 1 $^{0.33\%}$ & 0 $^{0.00\%}$ & 0 $^{0.00\%}$ & 0 $^{0.00\%}$ & 41 $^{6.83\%}$ & 0 $^{0.00\%}$ & 0 $^{0.00\%}$ & 0 $^{0.00\%}$\\ 
UbO & 15 $^{1.00\%}$ & 2 $^{0.67\%}$ & 0 $^{0.00\%}$ & 0 $^{0.00\%}$ & 0 $^{0.00\%}$ & 0 $^{0.00\%}$ & 76 $^{12.67\%}$ & 0 $^{0.00\%}$ & 0 $^{0.00\%}$ & 3 $^{0.3\%}$\\ 
PB & 18 $^{1.20\%}$ & 0 $^{0.00\%}$ & 0 $^{0.00\%}$ & 0 $^{0.00\%}$ & 1 $^{0.33\%}$ & 0 $^{0.00\%}$ & 37 $^{6.17\%}$ & 0 $^{0.00\%}$ & 0 $^{0.00\%}$ & 1 $^{0.1\%}$ \\ 

\midrule[1pt]
\end{tabular}
\end{table*}

\section{Limitations}
\label{sec:limitation}
Measuring web breakages due to adblocking is a hard problem due to the dynamic nature of the web and the limitations of the existing crawlers to imitate human behavior. Due to these reasons, researchers in the past have used different proxies to guesstimate the plausible impact of adblocking on the web. 

Websites contain a vast amount of HTML elements. It is nearly impossible to interact with every element present on the websites. Also, a lot of those elements are redundant and are unaffected by adblocking. Currently, it takes manual effort to visit the site and confirm that the elements in consideration are not working which takes significant time. To complete the measurements in time and cover a wide range of relevant elements, we focus on the subset of the popular breakage categories from Nisenoff et al.'s taxonomy. Although it does not cover the web breakage in its entirety, it provides a fair insight into the state of the breakages on the web. Also, due to the necessity of manual evaluation, our initial tests were conducted on a limited selection of websites. We aim to extend this examination to a broader range of sites in future research. 

To mitigate denial-of-service attacks, websites employ various techniques to protect them, such as automatic detection of crawlers and requiring users to solve CAPTCHAs to access the websites multiple times over a short duration. Both scenarios render assessment of the site infeasible. Also, a site might add extra noise in terms of server-side delays and return moderated content when visited via automated crawlers. This leads to inaccurate measurement data as well as a large chunk of websites becoming unavailable for testing.


A majority of web resources that the websites fetch are dynamic. They are either linked to the session or are fetched from servers with dynamic locations/IPs. For each measurement, we need to start a fresh session of the selenium as the elements can expire or become unresponsive with time within the same session. This leads to the inability to interact with different elements using source-based identifiers as their source location changes every time. Also, the dynamic behavior is reflected in the source code leading to difficulty in identifying the elements using XPaths. We, therefore, restrict our analysis to the web elements that are static over multiple crawls. Catapult tries to address these issues but has the inherent limitations that make it of limited use (Refer to Appendix \ref{sec:appendix}).


\section{Results}
\label{sec:results}
In this section, we highlight the results of the preliminary crawls that we did for each category. We aim to extend these crawls over the entire website testing pool set in future work. A subset of websites was chosen as a lot of these results require manual verification and it is very hard to do it on a large number of websites. Table \ref{tab:data} shows the corresponding data captured for each category.

\paragraph{Extension detection}
The number of websites that show a `disable adblock' prompt ABP, PB, and UbO are 25, 18, and 15 respectively. The number of websites detecting their presence would be greater than this but only these websites ask the users to disable the extension. The false positives obtained from the automated analysis were removed through manual analysis of the sites.

\paragraph{HTML elements}
Out of 300 websites we manually tested for the HTML we selected, only two sites show visible breakages in the buttons element for UbO. Out of them, only one site breaks for PB for the same button. One other site breaks in the drop-down element for PB. Both the sites work for ABP. One other site is redirected to a different website in the presence of ABP.

\paragraph{Resources not loaded}
As shown in Table \ref{tab:data}, out of the 1500 websites we tested, UbO has 76 sites with at least one image missing. ABP and PB have 41 and 37 sites with at least one image missing respectively.
We do find one website with missing video resources in the extension case but fail to catch it using Selenium as the URLs used to fetch the video are dynamic. Hence, it generates an XPath which is no longer part of the website's DOM tree.
An important point to note here is that the missing resources would not always result in visible breakages on the webpage. Frames hosting images can be of varying sizes and the chrome rendering engine can replace the missing images with the remaining image resources fetched by the website by shifting the page content upwards. We classify it as visible breakage since that particular resource is not present on the website anymore and hence cannot be located by a user looking for that specific element.

\paragraph{Browser/Page crash}
We do not find substantial evidence for the Browser/Page crashes where an adblocker prevents a website from loading due to the browser session being not present. 

\paragraph{Page unresponsiveness}
We qualify the page as unresponsive if it does not get loaded during the extension phase but gets loaded in both the non-extension phases i.e. before extension installation and after extension uninstallation. Based on our definition of unresponsiveness, we find that all the sites are responsive for ABP. We find three and one unresponsive sites for UbO and PB respectively.

Our findings reveal a critical insight: visible breakages due to adblockers are less common than anticipated. This suggests an increasing awareness among website developers about adblocking, leading to designs that maintain functionality despite it. However, the rise in anti-adblocking scripts underscores the importance of investigating invisible breakages~\cite{zhu18ndss}. These covert strategies, potentially acting as dark patterns, aim to subtly deteriorate the user experience, highlighting a need for further scrutiny in this area.



\section{Related Work}
Given the hardness of the problem of detecting visible breakages on websites due to adblocking, a limited number of researchers have attempted to address this problem. There have been a few studies that have attempted to measure the breakages using reasonable proxies to get a fair estimate of it. Amjad et al.~\cite{amjad} qualitatively studied web breakages on 383 websites by manually visiting them and identifying visible breakages. They measure breakages under three specific scenarios: 
control case (no adblocking), tracking and mixed JS blocking (with customized adblocker), and method-level JS blocking. The tracking and mixed JS blocking is closer to what we want to measure here. They find out that over 100 websites out of their pool experience some kind of visible breakage in this case.

Smith et al.~\cite{smith} developed a classifier to identify which filter list changes cause breakages on websites. To train their classifier, they used public GitHub issue reports of reported breakages due to filter list rules and recorded webpage behavior using PageGraph. Le~\cite{hieu} et al. also tried to measure breakages to verify if the automated filter rules generated by their tool are robust against website breakages. They observed that 4\% (10/272) of sites had breakages. They used the change in the amount of text and images on the webpage after applying filter list rules to approximate breakages on the webpages.

\section{Conclusion and Future Work}
\label{sec:discussion}

In this initial study, we investigate visible web breakages across five categories by examining live websites with the implementation of three different adblockers, each employing distinct adblocking technologies. This exploration offers insights into the capabilities and limitations of state-of-the-art web crawling tools for measuring such breakages, highlighting the complex nature of this issue and suggesting methods to mitigate it as much as possible. Although our research does not encompass all possible aspects of visible web breakages, it does provide valuable perspectives on the impact of adblockers on web functionality. Future endeavors will extend this analysis to a broader array of websites and delve into the phenomenon of invisible breakages.

Our approach marks a departure from previous methodologies that relied on proxies to estimate the extent of web breakages, thereby only being able to suggest an upper limit on their prevalence. By conducting live measurements, our work establishes a more concrete lower boundary for these disruptions. The findings reveal that visible breakages due to adblockers are less frequent than commonly perceived, challenging prevailing assumptions within the field. This investigation lays the groundwork for subsequent research, aiming to overcome current limitations and enhance the effectiveness of adblocking, thereby contributing to the advancement of a safer, more private online environment.

In this research, we focus on quantifying web breakages on live sites to gain a concrete understanding of how adblockers affect the web, offering a realistic lower estimate of the breakages occurring at any moment. However, our study does not encompass all types of breakages, particularly overlooking invisible breakages that will be addressed in future research. Invisible breakages often involve websites treating users with adblockers differently, leading to issues like delayed loading or reduced content quality, which worsen the user experience.
The significant discrepancy between the number of sites using adblocker detection scripts and those prompting users to disable adblockers underscores the importance of investigating how websites may be manipulating browser rendering processes to negatively impact users with adblockers. Our forthcoming studies aim to delve into these practices, examining whether websites force certain conditions that degrade the user experience for adblocker users.

Moreover, our manual site analyses have revealed discrepancies in behavior between automated crawling with Selenium in Xvfb mode and manual site visits, particularly regarding resource availability. This discrepancy suggests websites may respond differently to automated crawlers and web proxies, highlighting a critical area for future exploration. This finding points to the complex challenges in accurately measuring adblocker effects and the necessity for continued research in this area to better understand and mitigate web breakages.






%

\appendices
\section{Manual analysis of catapult tool}
\label{sec:appendix}
To study the robustness of the catapult tool, we record the landing pages of 200 websites and manually test them to check how they perform in the replay mode. We categorize them into four categories --- `no error' (websites showing no observable errors), `broken' (websites completely broken visibly with distorted layout), `slightly broken' (websites with a few missing elements with no distorted layout), and `unreachable' (websites that catapult could not record). We find that 112 websites fall into the `no error' category. These sites do not have a significant amount of ad resources on their pages. Nine websites fall into the broken category, 22 websites are slightly broken and 57 websites are unreachable. Slightly broken websites are high-content websites having multiple ad resources and other dynamically fetched animation/video web elements. A high number of unreachable sites highlights the instability of the catapult tool in recording them. Also, a large number of broken sites hosting ad content reflects the incapability of the tool to record and replay dynamically-fetched web elements.

\end{document}